\documentstyle[12pt,epsfig]{article}

\def\bc{\begin{center}}
\def\ec{\end{center}}
\def\be{\begin{equation}}
\def\ee{\end{equation}}
\def\bea{\begin{eqnarray}}
\def\eea{\end{eqnarray}}
\def\nn{\nonumber}

\newcommand{\lqcd}{\Lambda_{\rm QCD}}

\def\dirule{$\Delta I=1/2$ rule}
\def\oplus{O^{(+)}}
\def\ominus{O^{(-)}}
\def\oplmi{O^{(\pm)}}

\def\hat{\widehat}
\def\tilde{\widetilde}

\def\slash#1{\mbox{$\not \!\! #1$}}
\def\rDslash{{\overrightarrow{\slash D}}}
\def\lDslash{{\overleftarrow{\slash D}}}
\def\rdslash{{\overrightarrow{\slash \partial}}}
\def\ldslash{{\overleftarrow{\slash \partial}}}

\newcommand{\<}{\langle}
\renewcommand{\>}{\rangle}
\def\spose#1{\hbox to 0pt{#1\hss}}
\def\ltapprox{\mathrel{\spose{\lower 3pt\hbox{$\mathchar"218$}}
 \raise 2.0pt\hbox{$\mathchar"13C$}}}
\def\gtapprox{\mathrel{\spose{\lower 3pt\hbox{$\mathchar"218$}}
 \raise 2.0pt\hbox{$\mathchar"13E$}}}
\def\inapprox{\mathrel{\spose{\lower 3pt\hbox{$\mathchar"218$}}
 \raise 2.0pt\hbox{$\mathchar"232$}}}

\begin{document}

\pagestyle{empty} 
\begin{flushright}
EDINBURGH 97/6 \\
ROME1-1174/97 \\
ROM2F/97/25 \\
SHEP 97/15 \\
UW/PT 97-17 \\
\today
\end{flushright}
\centerline{\LARGE{\bf New lattice approaches to the $\Delta I=1/2$ rule}}
\vskip 0.3cm
\vskip 1cm
\centerline{\bf{C. Dawson$^a$,
G.~Martinelli$^b$, G.C.~Rossi$^c$, C.T.~Sachrajda$^a$,}}
\vskip 0.2cm
\centerline{\bf{S.~Sharpe$^{b,d}$,
M.~Talevi$^{e}$ and M.~Testa$^{b}$}}
\vskip 0.3cm
\centerline{$^a$ Dept. of Physics and Astronomy, University of Southampton,}
\centerline{Southampton SO17 1BJ, UK}
\smallskip
\centerline{$^b$ Dip. di Fisica, Univ. di Roma ``La Sapienza'' and
INFN, Sezione di Roma,}
\centerline{P.le A. Moro 2, I-00185 Roma, Italy}
\smallskip
\centerline{$^c$ Dip. di Fisica, Univ. di Roma ``Tor Vergata''
and INFN, Sezione di Roma II,}
\centerline{Via della Ricerca Scientifica 1, I-00133 Roma, Italy}
\smallskip
\centerline{$^d$ Physics Dept., University of Washington,
Seattle WA 98195, USA}
\smallskip
\centerline{$^e$ Department of Physics and Astronomy, University of Edinburgh,}
\centerline{The King's Buildings, Edinburgh EH9 3JZ, UK}

\vskip 2cm
\centerline{\bf ABSTRACT}
\begin{quote}

Lattice QCD should allow a derivation 
of the $\Delta I=1/2$ rule from first principles, 
but numerical calculations to date
have been plagued by a variety of problems.
After a brief review of these problems, we present several new methods
for calculating $K\to\pi\pi$ amplitudes.
These are designed for Wilson fermions, though they can be used also
with staggered fermions.
They all involve a non-perturbative determination of matching coefficients.
We show how problems of operator mixing can be greatly reduced by
using point-split hadronic currents,
and how CP violating parts of the $K\to\pi\pi$ amplitudes can be
calculated by introducing a fake top quark.
Many of the methods can also be applied to the calculation of two body
non-leptonic $B$-meson decays.

\end{quote}
\vfill

\newpage

\pagestyle{plain}
\setcounter{page}{1}

\section{Introduction}

One of the least well understood features of hadronic physics is the
$\Delta I=1/2$ rule in non-leptonic kaon decays.
Decays in which isospin changes by $\Delta I=1/2$ are greatly enhanced over
those with $\Delta I=3/2$.
The particular example we focus on here is the ratio of
amplitudes for $K\to\pi\pi$ decays:
\begin{equation}
{{\cal A}(K\to\pi\pi[I=0]) \over {\cal A}(K\to\pi\pi[I=2]) } \approx 22 \,.
\label{eq:dirule}
\end{equation}
Although the origin of this large enhancement is not well understood,
we do know that, in a QCD-based explanation,
most of the enhancement must come from long distance, 
non-perturbative physics.
This is because the contribution from scales where perturbation theory is
reliable, say $p \gtapprox 2\;$GeV,
is known to enhance the $\Delta I=1/2$ amplitude by only a factor of
about two.
Attempts to understand the remainder of the enhancement using models of
non-perturbative physics have had partial success \cite{bbg},
but we are far from having a convincing demonstration that QCD does explain
the \dirule.

In principle, lattice QCD is well suited to study this issue~\cite{acient}.
Indeed, 
practical approaches have been developed for both Wilson \cite{MMRT,direct}
and staggered fermions \cite{toolkit,sharpepatel}.
The problem is that these methods have not yet yielded useful results.
Indirect methods, based on using $K\to\pi$ and $K\to$vacuum amplitudes, have
large statistical errors \cite{GMlat89,SSlat90},
while direct calculations of $K\to\pi\pi$ amplitudes
have been done at unphysically large quark masses,
for which the presence of a scalar resonance may
distort the two pion signal~\cite{GMlat89,BSlat89,gavela}.
There has been little work on the problem in recent years.

In this paper we revisit the lattice approach for the case of Wilson
fermions (or $O(a)$ improved versions thereof).  We reevaluate the
existing methods, and propose a variety of new approaches.  These vary
from a method of comparable simplicity to that of
Bernard {\em et al.}~\cite{direct}, to more speculative ideas that are
likely to require much smaller lattice spacings than those presently
available.  The latter are needed to calculate the CP violating parts
of the $K\to\pi\pi$ amplitudes. We also reappraise the calculation of
$K\to\pi$ matrix elements in the light of the recent progress made in
non-perturbative renormalization techniques.

Existing methods rely on lowest order chiral perturbation theory to
connect the unphysical amplitudes which are calculated on the lattice
to the desired physical amplitude. One advantage of some of our new
methods is that they are not dependent on chiral perturbation theory.
This means that they can, in principle, also be used to study $B$-meson
decays.

The outline of this paper is as follows.  In the next section we
summarise the problems faced by lattice calculations, and then, in
sec.~\ref{sec:direct}, recall how these are avoided by the direct
method of Bernard {\em el al.} \cite{direct}. With the stage thus set,
in sec.~\ref{sec:new} we then present the first three of our new
methods, which also involve the calculation of $K\to\pi\pi$
amplitudes. This is followed in sec.~\ref{sec:pp} by a reappraisal of
the indirect method using $K\to\pi$ amplitudes. In
sec.~\ref{sec:jj}, we present a more speculative method for studying
the $\Delta I=1/2$ rule, which is based on the short distance expansion
of the $T$-product of two weak currents.
Finally in sec.~\ref{sec:top} we suggest a possible
idea for a non-perturbative determination of the CP-violating part of
the $\Delta S=1$ weak Hamiltonian by introducing a fictitious top
quark. Section~\ref{sec:concs} contains our conclusions.

\section{The problem}\label{sec:problem}

In this section we briefly review the source of the difficulty in
calculating ${\cal A}(K\to\pi\pi)$ with Wilson fermions.  Further
details can be found in refs.~\cite{MMRT,BernardTasi} and references
therein.

For scales below $M_{_W}$, but above the charm quark mass, the $\Delta S=1$
part of the effective weak Hamiltonian can be written as 
\bea
{\cal H}_{\rm eff}^{\Delta S=1}
&=&
\lambda_u {G_F \over \sqrt2} 
\left[ C_+(\mu, M_{_W}) \oplus(\mu) + C_-(\mu,M_{_W}) \ominus(\mu) \right]\,,
\label{eq:HW}\\
\oplmi &=& \left[ (\bar s \gamma_\mu^L d)(\bar u \gamma_\mu^L u)
 \pm (\bar s \gamma_\mu^L u)(\bar u \gamma_\mu^L d) \right]
  -  \left[ u \leftrightarrow c \right] \,,
\label{eq:oplmidef}
\eea
where $\gamma_\mu^L=\gamma_\mu (1\!-\!\gamma_5)/2$ and 
$\lambda_u=V_{ud}V_{us}^*$.
Here and in the following we use the Euclidean metric.
We are ignoring for the moment the contribution which arises when
the top quark is integrated out. This is suppressed by $\lambda_t/\lambda_u$,
where $\lambda_t=V_{td}V_{ts}^*$, and, 
in the CP conserving sector,  
makes a very small contribution to the decay amplitudes.

The operators $\oplmi$ have different transformation properties under isospin.
In particular, $\ominus$ is pure $I=1/2$,
whereas $\oplus$ contains parts having both $I=1/2$ and $I=3/2$.
An explanation of the \dirule\ thus requires
that the $K\to\pi\pi$ matrix element of $C_- \ominus$ 
be substantially enhanced compared  to that of $C_+ \oplus$.
The short distance, perturbative part of the enhancement is that provided by
the ratio of Wilson coefficients, $C_-/C_+$,
while the long distance, non-perturbative contribution comes from
the ratio of matrix elements of the operators.
Consider first the short distance contribution.
At $\mu=M_{_W}$, the two Wilson coefficients have nearly equal magnitudes, 
$|C_-/C_+|=1 + O[\alpha_s(M_{_W})]$.
The enhancement arises from the renormalization group evolution
down to $\mu \sim 2\;$GeV, at which scale one finds $|C_-/C_+| \approx 2$.
This factor is too small by
an order of magnitude to explain the \dirule.

\begin{figure}[t] 
\vspace{0.1cm}
\centerline{\epsfig{figure=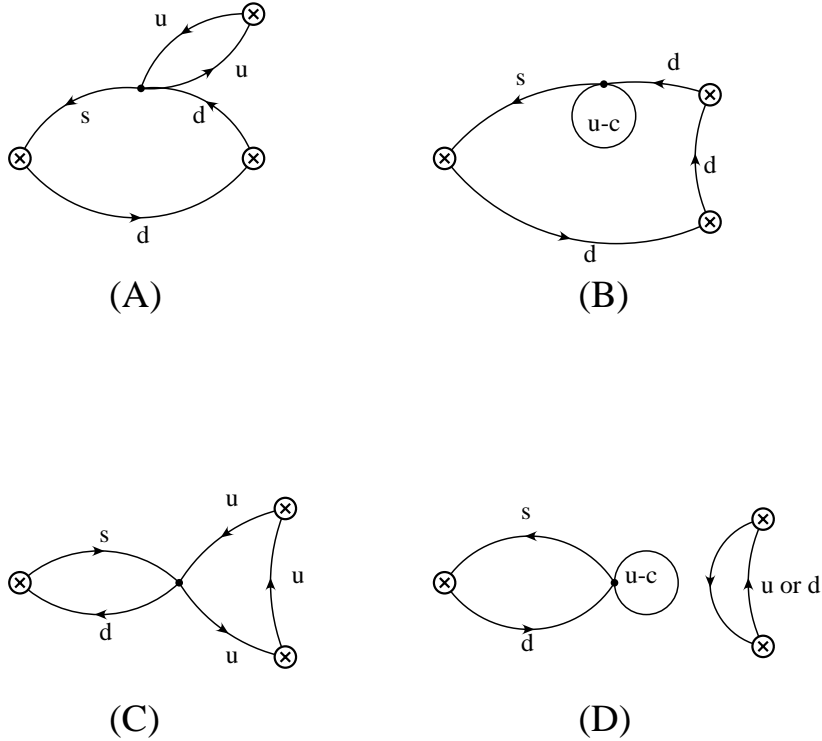,height=10cm}}
\caption{{\it Contractions contributing to the $\Delta I=1/2$ amplitude.
The lines represent propagators
on a background gauge field, and the resulting contraction is to be
averaged over gauge configurations with the appropriate measure.
The dot represents the operators $\oplmi$. Different colour contractions
are not distinguished.}}
\label{fig:wick}
\end{figure}

The remainder of the enhancement must come from the matrix elements of
the operators, and these are the quantities that we wish to calculate
on the lattice.  In fig.~\ref{fig:wick} we show the Wick contractions
that contribute to such matrix elements.  The only part of ${\cal
H}_{\rm eff}^{\Delta S=1}$ which gives rise to $\Delta I=3/2$
transitions is the $I=3/2$ part of $\oplus$, and for this operator
only contractions (A) and (C) contribute.  In contrast, all four
contractions are non-vanishing for the $I=1/2$ parts of $\oplmi$.
Thus, in order to reproduce the \dirule, the sum of the 
contributions of (B) and
(D) must be an order of magnitude larger than those of (A) and (C).
In fact, as discussed below, contractions (C) and (D) give non-leading
contributions in chiral perturbation theory compared to (A) and (B).
Thus we expect contraction (B) to be the ``source'' of the \dirule.

There are two major difficulties which arise in the calculation 
of these diagrams using lattice QCD. 
\begin{enumerate}
\item
Decay amplitudes into two or more particles cannot be calculated
directly in Euclidean space.
This follows from the theorem of Maiani and Testa \cite{MT}.
One must, instead, use a model to analytically continue 
correlations functions from Euclidean to physical momenta.
The only exception is for final state particles at rest relative to
each other, 
in which case there is no phase generated by final state interactions.
This problem afflicts both $\Delta I=1/2$ and $3/2$ amplitudes,
although it is likely to be worse for the former since final state
interactions are stronger for two pions having $I=0$ \cite{isgur}. 
\item
The ``penguin'' diagrams (B) and (D) allow the
$I=1/2$ operators to mix with operators of lower dimension with coefficients which diverge as inverse powers of the lattice spacing.
This mixing leads to contributions to the amplitudes which
are lattice artifacts, and must be subtracted.
One must also account for the mixing with other
operators of dimension six, although this is less difficult because
the mixing coefficients do not diverge in the continuum limit.
\end{enumerate}
In the remainder of this section we expand upon the latter problem. 
We consider only the negative parity parts of $\oplmi$, since the parts
with positive parity do not contribute to $K\to\pi\pi$ amplitudes.

Lower dimension operators which can appear on the lattice
must have the same flavour and CPS parity as $\oplmi$.
CPS is the transformation obtained by combining
CP with $(s\leftrightarrow d)$ interchange \cite{politzer}.
Both $\oplmi$ have positive CPS parity.
These symmetries allow mixing with only two lower dimension operators
(aside from operators which vanish by the equations of motion):
\bea
O_p &=& (m_s-m_d) \bar s \gamma_5 d \,, 
\label{eq:opdef}\\
\widetilde{O}_\sigma &=& 
g_0 (m_s-m_d) \bar s \sigma_{\mu\nu} \tilde G_{\mu\nu} d \,,
\eea
Here $g_0$ is the bare coupling constant
and $\tilde G_{\mu\nu}$ the dual field strength. 
The factors of $m_s\!-\!m_d$ are required by CPS symmetry.
At dimension six, CPS forbids mixing with other four-fermion operators,
and SU(4) flavour forbids mixing between $\oplus$ and $\ominus$.
It is noteworthy that the allowed mixing for the negative parity
parts of $\oplmi$ is, up to this stage, exactly as in the continuum,
despite the fact that chiral symmetry is explicitly broken
on the lattice \cite{MMRT}.

The mixing with $O_p$ and $\widetilde{O}_\sigma$ 
is further constrained by the GIM
mechanism, which implies that
the mixing coefficients vanish identically when $m_c=m_u$.
In continuum perturbation theory, chiral symmetry 
requires that the mixing coefficients are quadratic functions of the
quark masses and thus vanish as $m_c^2-m_u^2$.
On the lattice, by contrast, chiral symmetry is broken, and
the GIM cancellation gives rise only to the factor $m_c-m_u$.
This additional factor is sufficient, however,
to make the mixing of $\oplmi$ with $\widetilde{O}_\sigma$ an effect of $O(a)$.
Since in this work we are not attempting to remove $O(a)$ effects
from the matrix elements of $\oplmi$,
we do not need to consider mixing with $\widetilde{O}_\sigma$.
We therefore arrive at the following form of the renormalized operator 
\be 
O^{(\pm)}(\mu)= Z^{(\pm)}(\mu a, g^2_0) \left[ O^{(\pm)}(a)
+ (m_c -m_u)  \frac{C^{(\pm)}_p}{a} O_p(a) \right]  + O(a)\,.
\label{eq:rops} 
\ee Here $O^{(\pm)}(a)$ and $O_p(a)$ are bare lattice operators,
$C^{(\pm)}_p$ are the mixing coefficients, and $Z^{(\pm)}$ are the
renormalization constants which cancel the logarithmic divergence of
the bare four-fermion operator.  The precise definition of the lattice
quark masses in eq.~(\ref{eq:rops}) is unimportant, since in all
practical methods the entire coefficient of $O_p$ is determined
non-perturbatively.

We now show that, in spite of the fact that it multiplies a linear
divergence (see eq.~(\ref{eq:rops})), it is sufficient to determine
$C^{(\pm)}_p$ with an error of $O(a)$, to obtain the physical
amplitudes to the same precision. We first note that, in the
continuum, $O_p$ is proportional to the divergence of the axial
current, apart from terms which vanish by the equations of motion.
Thus, in the continuum, it does not contribute to on-shell matrix
elements for which the momentum inserted by the operator, $\Delta p$,
vanishes.
This is why the mixing of $\oplmi$
with $O_p$ in the continuum is irrelevant to physical amplitudes.  On
the lattice, however, the breaking of chiral symmetry leads to $O(a)$
corrections to the PCAC equation even after renormalization~\cite{boc}:
\be
\langle h_1|\partial_\mu A_\mu - (m_s+m_d) P|h_2\rangle
= \langle h_1| \bar X_A |h_2 \rangle = O(a) \,.
\label{eq:PCAClat}
\ee
Here $A_\mu=Z_A \bar s \gamma_\mu \gamma_5 d$
and $P=Z_P \bar s\gamma_5 d$ are the renormalized operators,
$m_d$ and $m_s$ are renormalized quark masses,
$\partial_\mu$ is a lattice derivative,
and $h_{1,2}$ represent hadronic states.
Thus the $K\to\pi\pi$ matrix element of the subtraction term is
\be  
\frac{C_p^{(\pm)}}{a} (m_c -m_u) (m_s - m_d) 
\langle \pi \pi \vert \bar  s \gamma_5 d \vert K \rangle =
\frac{C_p^{(\pm)}}{a} (m_c -m_u) \frac{m_s - m_d}{m_s + m_d} {1\over Z_P} 
\langle \pi \pi \vert \partial_\mu A_\mu - \bar X_A \vert K \rangle \,.
\label{eq:problem}
\ee 
The $\partial_\mu A_\mu$ term leads to the divergent contribution
proportional to $\Delta p/a$.  But when $\Delta p=0$, the term
proportional to $\bar X_A/a$ remains, and is of $O(1)$ up to logarithms.
Thus the subtraction is still necessary, but the magnitude of the subtraction
does not diverge in the continuum limit, and implies that it is sufficient to
know $C_p^{(\pm)}$ with an error of $O(a)$.

In summary, to calculate the physical $K\to\pi\pi$ matrix elements of
$\oplmi$ one needs both a model to do the analytic continuation from
Euclidean to Minkowski space, and a method to subtract the operator
$(m_c-m_u) O_p$ with appropriate coefficients $C_p^{(\pm)}$. Having
summarised the problem, the remainder of the paper is devoted to
illustrate several methods which can be used to extract the physical
amplitudes.

\section{Calculations with ${\mathbf m_s=m_d.}$}\label{sec:direct}

Bernard {\em et al.} proposed an ingenious solution to both problems 
\cite{direct}.
They suggest working with $m_s=m_d$, and calculating the 
Euclidean amplitude in which all three particles are at rest:
\be
{\cal A}_{\rm m_s=m_d}^{(\pm)} = \langle \pi(\vec p_1\!=\!0) \pi(\vec p_2\!=\!0)  
| O^{(\pm)}(\mu) | K(\vec p_K\!=\!0) \rangle\bigg|_{m_s=m_d} \,.
\label{eq:b+s}
\ee
Setting $m_s=m_d$ causes $O_p$ to vanish identically, 
and so removes the need for the subtraction.
Working with the two pions at rest solves the problem of final
state interactions.
The final step in the method is to extrapolate from the
unphysical amplitudes ${\cal A}_{\rm dir}^{(\pm)}$
to the physical amplitudes ${\cal A}_{\rm phys}^{(\pm)}$
using lowest order chiral perturbation theory.
The result is that
\be
{\cal A}_{\rm phys}^{(\pm)} = { m_{K,\rm phys}^2 - m_{\pi,\rm phys}^2 \over 
2 m_{K,\rm lat}^2}\ {\cal A}_{\rm m_s=m_d}^{(\pm)} 
\left[ 1 + O\left(m_K^2 \over \Lambda_\chi^2\right) \right] \,,
\label{eq:direct}
\ee
where $m_{K,\rm lat}$ is the mass of the lattice ``kaon'' used in
calculating ${\cal A}_{\rm m_s=m_d}^{(\pm)}$.
Due to the choice $m_s=m_d$, this is also the mass of the lattice ``pion''.
$\Lambda_\chi\approx 4 \pi f_\pi$ is the scale which determines
the size of higher order terms in chiral perturbation theory.
Such terms typically give $25\%$ corrections if 
$m_{K,\rm lat} \approx m_{K,\rm phys}$.
The direct method should thus be adequate to test the \dirule\
at a semiquantitative level.

The main advantage of this method is its simplicity.  It determines
the coefficient of the contribution to ${\cal A}_{\rm phys}^{(\pm)}$
of leading order in chiral perturbation theory without the need for a
subtraction.  A further simplification is that one only needs to
calculate the contractions (A) and (B) in fig.~\ref{fig:wick},
because, due to CPS symmetry, the other two contractions vanish
identically when $m_s=m_d$.  In fact, since one can show that the
physical amplitude depends on quark masses only at non-leading order
in chiral perturbation theory, the contributions of diagrams (C) and
(D), which are proportional to $m_s-m_d$, must be of non-leading
order.

The main disadvantage of the method is its dependence on chiral
perturbation theory. Such dependence is unavoidable once one has
set $m_s=m_d$. Of particular concern is the fact that
the final state interaction phase in the $I=0$ two pion channel is
substantial at the physical point $s=m_K^2$ (where $s$ is the 
square of the two-pion centre-of-mass energy), even though it is
a non-leading effect in chiral perturbation theory.
Thus it is possible that the corrections in eq.~(\ref{eq:direct})
are larger than the estimate $m_K^2 / \Lambda_\chi^2$ 
\cite{isgur}.
In principle one could reduce this error by using non-leading order
chiral perturbation theory.
This, however, requires knowledge of coefficients which are unavailable
from experiment \cite{kambor}.

\section{Alternative methods}
\label{sec:new}

We wish to develop methods which allow us to reduce, and ultimately
remove, the dependence on chiral perturbation theory.
To do this we must use quark masses which are closer to their physical
values, and in particular we must consider $m_s > m_d$.
Working with non-degenerate quarks implies that
the mixing with $O_p$ is present, and must
either be subtracted, or shown to be unimportant.
We have devised a number of new methods which accomplish this goal.
In this section we describe, in order of increasing complexity,
the three which require relatively small changes in methodology
compared to the proposal of section~\ref{sec:direct}.
More speculative methods are described in subsequent sections.

Before describing our methods, we first comment on the the overall
renormalization factors $Z^{(\pm)}(\mu a, g_0^2)$.  These must be
calculated for all the methods described in this section and in
sections~\ref{sec:direct} and ~\ref{sec:pp}.  In the past
$Z^{(\pm)}$ have been determined using one-loop perturbation
theory.  Although this may be satisfactory for a test
of the \dirule, a non-perturbative determination is clearly
preferable.  We wish to point out that such a determination can be
made by a straightforward extension of the non-perturbative methods
used to normalize $\Delta S=2$ operators using quark
states~\cite{talevi}.  The key observation is that the anomalous
dimensions of $\oplmi$ are unchanged by mixing with the pseudoscalar
density, and thus coincide with those of the operators
\be 
\bar O^{(\pm)}= \Bigl(\bar \psi_1 \gamma_\mu  \psi_2 \, \bar\psi_3
\gamma_\mu \gamma_5 \psi_4 + 
\bar \psi_1 \gamma_\mu \gamma_5 \psi_2 \, \bar\psi_3
\gamma_\mu  \psi_4 \Bigr) \pm \Bigl( 2 \leftrightarrow 4 \Bigr) \, ,
\ee
where the subscripts $1,2,3,4$ label four different quark flavours.
Only open current-current diagrams need be calculated; penguin-type
diagrams with quark loops do not contribute.  The non-perturbative
calculation of the renormalized $\bar O^{(\pm)}$ is currently
underway~\cite{contil}.

\subsection{Method 1}

We first describe the method and then explain why it is valid.
The ingredients are
\begin{itemize}
\item Work with a non-perturbatively $O(a)$ improved fermion action,
for which there are no errors of $O(a)$ in the spectrum, and the
on-shell amplitudes of the improved currents obey the continuum chiral
Ward identities up to $O(a^2)$. This can be accomplished using the
methods of ref.~\cite{luscher}.
\item
Use the lattice four-fermion operator, suitably normalized,
but do not subtract the $O_p$ term. In other words, set $C_p^{(\pm)}=0$.
\item
Choose quark masses such that $m_K= 2m_\pi$.
\item
Calculate the $K\to\pi\pi$ amplitudes with all particles at rest.
Given the choice of meson masses this means that $\Delta p=0$. 
The resulting amplitudes are denoted ${\cal A}_{m_K=2m_\pi}^{(\pm)}$.
\item
Determine the physical amplitudes
using lowest order chiral perturbation theory, which gives
\be
{\cal A}_{\rm phys}^{(\pm)} = { m_{K,\rm phys}^2 - m_{\pi,\rm phys}^2 \over 
 m_{K,\rm lat}^2 - m_{\pi,\rm lat}^2}\, {\cal A}_{m_K=2m_\pi}^{(\pm)} 
\left[ 1 + O\left(m_K^2 \over \Lambda_\chi^2\right) \right] \,.
\label{eq:N1}
\ee
\end{itemize}

The validity of this method is based on the following observations. 
If one uses fully $O(a)$ improved fermions,
and the fully $O(a)$ improved axial current, $A_\mu^I$,
and pseudoscalar density, $P^I$,
then the corrections to the PCAC equation (\ref{eq:PCAClat}) 
are of $O(a^2)$ rather than of $O(a)$ \cite{luscher}.
Furthermore, the improved pseudoscalar density is simply proportional to the
bare lattice operator $P^I=Z_P^I \bar s\gamma_5 d$.
Thus the argument following eq.~(\ref{eq:PCAClat}) now implies that
the matrix element of the subtraction term is
\be  
\frac{C_p^{(\pm)}}{a} (m_c -m_u) (m_s - m_d) \langle \pi \pi
\vert \bar  s \gamma_5 d \vert K \rangle =
\frac{C_p^{(\pm)}}{a} (m_c -m_u) \frac{m_s - m_d}{m_s + m_d} {1\over Z_P^I} 
\langle \pi \pi \vert \partial_\mu A_\mu^I \vert K \rangle 
+ O(a) \,.
\label{eq:soln1}
\ee 
The important point is that the discretization error on the r.h.s. of
this result is now of $O(a)$ rather than of $O(1)$ 
[cf. eq.~(\ref{eq:problem})].
This means that if we can set $\Delta p=0$, 
so that the term involving $\partial_\mu A_\mu^I$ vanishes,
the subtraction leads to corrections only of $O(a)$.
Since these are of the size that we are neglecting,
it is not necessary to do the subtraction.
Note that to make this argument,
we do not need to know the form of $A_\mu^I$, nor the
value of the renormalization constants $Z_A^I$ and $Z_P^I$.

Since we have chosen quark masses such that $m_K=2 m_\pi$,
and chosen all particles to have $\vec p=0$, 
we do have $\Delta p=0$. 
Furthermore, the use of pions at rest allows us to avoid final state 
interactions, as in the $m_s=m_d$ method.

We view this new method as complementary to the method of
section~\ref{sec:direct}. It is only slightly more difficult to
implement practically.  One complication is the need to use fully
$O(a)$ improved Wilson fermions.  This is, however, becoming standard
in numerical simulations, now that the necessary ``clover''
coefficient $c_{SW}$ has been determined non-perturbatively for
quenched QCD \cite{luscher}. Note that the method of
section~\ref{sec:direct} can be used equally well with improved
fermions, but that this does not provide any particular advantage over
unimproved fermions. The results for $K\to\pi\pi$ amplitudes will have
errors of $O(a)$ in both cases.  To remove these errors would require
not only improving the action but also improving the operators
$\oplmi$.

A second potential technical complication is the need to include all
four Wick contractions of fig.~\ref{fig:wick}.  One can, however,
consistently drop diagrams (C) and (D) since they are non-leading in
chiral perturbation theory, while we are relying in eq.~(\ref{eq:N1})
on the leading order term.

A possible advantage of this method is that it requires a smaller
extrapolation to the physical point, since one is using quark masses
which are closer to those of the physical quarks.  Furthermore, if one
includes diagrams (C) and (D), one can perform a partial test of the
convergence of the chiral expansion.  Diagram (C) is straightforward
to calculate, while the disconnected diagram (D) presents more
technical difficulties.  We expect, however, that since (D) is Zweig
forbidden, its contribution will be smaller than that of (C).

\subsection{Method 2}

A potential problem with the previous method is that the momentum
inserted will not, in practice, be exactly zero.
If so, the part of the $O_p$ term proportional to $\partial_\mu A_\mu^I$
will give a ``small'' but divergent contribution proportional to $\Delta E/a$,
where $\Delta E=m_K- 2 m_\pi$.
We can mitigate this potential problem by determining $C_p^{(\pm)}$ 
non-perturbatively. We propose to do so by applying the condition
\be
\langle 0 \vert \oplmi(\mu) \vert K \rangle = 0 \,.
\label{eq:subt}
\ee
We then proceed as in the previous method, 
except now keeping the subtraction term.
Note that implementation of this condition does not require knowledge
of the overall normalizations $Z^{(\pm)}$.

For this relatively small increase in effort we effectively
determine the $O(1)$ part of $C_p^{(\pm)}$, 
and thus reduce the error from the incomplete subtraction of the $O_p$ term 
by one power of $a$.
In other words, whereas in Method 1 ignoring the subtraction term
leads to an error $\sim \Delta E/a + O(a)$,
in Method 2 the residue after implementing eq.~(\ref{eq:subt}) is
$\sim \Delta E + O(a^2)$.
Since there are other sources of $O(a)$ errors,
the total error in the $K\to\pi\pi$ matrix element is $\sim\Delta E + O(a)$.
Thus the extrapolation to $\Delta E=0$ will be considerably less delicate.
Furthermore, since the method no longer relies on the improved PCAC relation,
it can be used with unimproved, or tree-level improved, Wilson fermions. 

We now explain how the condition (\ref{eq:subt}) is justified.
At leading order in chiral perturbation theory,
the $K\to$vacuum matrix element takes the form~\cite{politzer} 
($f_\pi \sim 132$ MeV)
\begin{equation} 
\<0\vert  O^{(\pm)}(\mu) \vert K^0\> = i\,\delta^{(\pm)}\,
\frac{m^2_K - m^2_\pi}{f_\pi} \,.
\label{eq:SPT1}
\end{equation}
The coefficients $\delta^{(\pm)}$ do not contribute to the
physical $K\to\pi\pi$ matrix element (see eq.~(\ref{eq:SPT3}) below),
and thus can be set to zero.
A similar unphysical arbitrariness remains away from the chiral limit.
The point is that we can always redefine the renormalized operators
as follows~\cite{pate} 
\be
\oplmi \longrightarrow \oplmi
+ F^{(\pm)} (m_c-m_u) O_p \,.
\label{eq:ambig}
\ee
where the finite coefficients $F^{(\pm)}$ are of $O(\lqcd)$ but
otherwise arbitrary.
This redefinition does not change the physical matrix elements of 
$\oplmi$ since $O_p$ is a total divergence.
But it does affect matrix elements in which momentum is inserted,
such as those appearing in the condition of eq.~(\ref{eq:subt}).
This condition can thus be fulfilled by an 
appropriate choice of $F^{(\pm)}$.
In particular, it can also be used if the strange quark is
replaced by the bottom quark.

The possibility of redefining the renormalized operators by finite
terms, as in eq.~(\ref{eq:ambig}), implies that dependence of the
$K\to\pi\pi$ matrix elements on $\Delta E$ is of $O(1)$ and does not
vanish with the lattice spacing as one would na\" ively expect.

\subsection{Method 3}

The previous methods require the use of quark masses
whose ratio $m_s/m_d$ differs from its physical value.
To move closer to physical quark masses while holding $\Delta E=0$ 
we must work with final states in which $\vec p_\pi \ne 0$. 
This forces us to deal with the Maiani-Testa theorem.
We propose doing so using the method of Ciuchini {\em et al.} (CFMS)
\cite{cfms}. Given an analytic parameterization 
of the two pion scattering amplitude
(e.g. resonance dominance) CFMS show how, in principle, one can extract
the magnitude and phase of the physical amplitude by studying the Euclidean
amplitude as a function of time for a variety of final pion momenta.

An assumption made by CFMS is the ``smoothness" of the off-shell amplitudes.
In the absence of resonances coupled to the final-state mesons, this
requires that the $K\to\pi\pi$ amplitudes do not vary rapidly with $\Delta p$.
If there is a nearby resonance, CFMS assume that it dominates
the momentum dependence of the amplitudes. 
The ``smoothness" hypothesis is, in this case,  the assumption
that the couplings of the resonance to the final-state particles
are  smooth functions of the external momenta. 
It is then is possible to find a simple parametrization
of  the amplitudes which describes their rapid variation
(due to the presence of the resonance) with $\Delta p$.

Thus we propose using the method of CFMS after fixing $C_p^{(\pm)}$ with
the conditions eq.~(\ref{eq:subt}).
This removes the terms proportional to $\Delta p/a$ which might violate
the smoothness hypothesis.
Since this method makes no use of chiral perturbation theory,
it can be applied also $B$ decays.
The only restriction is that the GIM mechanism must be operative,
which limits one to processes which do not involve top quark loops.

\section{${\mathbf K\to\pi}$ Matrix Elements}
\label{sec:pp}

In this section we reconsider the method suggested in ref.~\cite{MMRT}
which uses the $K\to\pi$ matrix elements of the positive parity part
of the weak Hamiltonian.  This method relies on chiral perturbation
theory, and in this respect we expect it to be of comparable accuracy
to the direct method of Bernard {\em et al.} and the first two of new
methods introduced in the previous section. 
Since only single-particle states are involved, the
advantage of the ``$K\to\pi$ method'' is that it is technically
easier to extract the relevant matrix elements.
The disadvantage is that the operator mixing problem
is much more complicated than for the negative parity part of ${\cal
H}_{\rm eff}^{W}$, making an accurate evaluation of the matrix element
of the renormalized operator difficult.

We begin by recalling how the physical amplitude is obtained from the
knowledge of the properties of the $K\to\pi$ amplitude exploiting the
Soft-Pion Theorems (SPTs).  At leading
order in chiral perturbation theory the physical amplitude takes the
form (for $\Delta p=0$)
\be
\<\pi^+\pi^-\vert  O^{(\pm)}(\mu)\vert K^0\> =
i\,\gamma^{(\pm)}\, {m^2_K - m^2_\pi\over f_\pi} 
\,. \label{eq:SPT3}
\ee
The coefficients $\gamma^{(\pm)}$ appear also in the expression for
the $K\to\pi$ matrix element 
\begin{equation}
\<\pi^+(p)\vert  O^{(\pm)}(\mu) \vert K^+(q)\> =
-\delta^{(\pm)}\, {m^2_K\over f_\pi^2} + \gamma^{(\pm)}\,p\cdot q
\,.\label{eq:SPT2}
\end{equation}
By studying this matrix element as a function of $p\cdot q$ one can,
in principle, determine $\gamma^{(\pm)}$, from which we obtain the
$K\to\pi\pi$ matrix elements up to corrections of order
$m_K^2/\Lambda_\chi^2$.

The expression for the positive-parity components of the
renormalized $\oplmi$ is \cite{MMRT}~\footnote{Although, for
convenience, we use the same symbol, $Z^{(\pm)}$, the values of the
overall renormalization constant for the positive parity part is in
general different from that for the parity violating part, as a result
of the explicit chiral symmetry breaking on the lattice. The operators
$O^{(\pm)}$ in this section also refer to their positive parity
components.}  
\bea 
\oplmi(\mu) &\equiv & Z^{(\pm)}(\mu a,
g_0^2)\,O^{(\pm)}_{\mathrm sub} \nonumber\\ & = & Z^{(\pm)}(\mu a,
g_0^2) \left[\oplmi(a) + \sum_{i=1}^4 C_i^{(\pm)} \oplmi_i(a) \right.
\nonumber\\ &&\left.  +(m_c-m_u) C^{(\pm)}_\sigma O_\sigma(a)
+(m_c-m_u) {C^{(\pm)}_s \over a^2} O_s(a) \right] \,.
\label{eq:pcmixing} \eea 
Here $\oplmi_i$ are four-fermion operators of dimension 6
which have different chirality from $\oplmi$. They are listed in
refs. \cite{BernardTasi}, \cite{talevi}--\cite{jlqcd}.  The remaining
operators are of lower dimension 
\bea O_\sigma &=& g_0 \bar s
\sigma_{\mu\nu} G_{\mu\nu} d \,,\label{eq:osigmadef}\\ O_s &=& \bar s
d \,. 
\label{eq:osdef}\eea 
Comparing the result in eq.~(\ref{eq:pcmixing}) with that for the
negative parity parts of $\oplmi$, eq.~(\ref{eq:rops}), we see that
CPS symmetry provides much weaker constraints for the parity
conserving parts. In particular, the loss of the factor of $m_s-m_d$
means that the mixing with the scalar density diverges like $1/a^2$.

Reference~\cite{MMRT} suggested the following approach for determining
the mixing coefficients: use perturbative values for the coefficients
of the dimension 6 operators and $O_\sigma$, but determine
$C^{(\pm)}_s$ non-perturbatively.  The latter determination is to be
made by adjusting $C^{(\pm)}_s$ until the momentum independent part of
the $K\to\pi$ matrix element in eq.~(\ref{eq:SPT2}) vanishes,
i.e. $\delta^{(\pm)}=0$. This is the positive parity analogue of the
condition~(\ref{eq:subt}), and can be justified by similar arguments.

One problem with this approach is that a perturbative determination of
mixing coefficients is often not reliable.  This is exemplified by the
case of $\Delta S=2$ operators, where the operator with
non-perturbatively determined coefficients has, to a very good
approximation, the expected chiral behaviour, while that with
perturbative coefficients does not \cite{talevi,jlqcd}. Nevertheless,
this approach should be pursued, because it could provide 
semiquantitative results for the \dirule. To date, no useful results
have been obtained.

There are three approaches which have been proposed to determine the
subtraction coefficients of four-fermion operators in a
non-perturbative way: Gauge Invariant Ward Identities
(GIWIs)~\cite{MMRT}, Ward Identities on quark states~\cite{jlqcd} 
 and non-perturbative renormalization between quark
states at large external momenta~\cite{talevi}. 

We start with a discussion of the GIWIs, reformulated in the light of
the recent progress in the exploitation of Ward identities in the 
context of improved actions~\cite{luscher}. As we shall explain below,
this is the most promising approach in the presence of mixing with
lower dimensional operators (and is more general than the present case).
Consider the following Ward identity in the chiral limit:
\begin{equation}
\sum_{y \in \cal{R}} \langle \partial_\mu A^f_\mu(y)\,
O^{(\pm)}_{\mathrm sub}(0)\, \Phi(x_1, x_2, \cdots, x_n)\rangle = -i
\langle\frac{\delta O^{(\pm)}_{\mathrm sub}(0)}{\delta\alpha^f}\,
\Phi(x_1, x_2, \cdots, x_n)\rangle \ ,
\label{eq:giwi}\end{equation}
where ${\cal R}$ is a region of space-time containing the origin
bounded by two hyperplanes $y_4=-t_{a}$ and $y_4=t_{b}$ and $f$ labels the
flavour component of the axial transformation.
$\Phi$ represents
a multilocal gauge-invariant operator, with $x_1, x_2, \cdots,x_n$ all
lying outside ${\cal R}$.  $\delta O^{(\pm)}_{\mathrm
sub}/\delta\alpha^f$ denotes the variation of the operators under
infinitesimal axial rotations of the fields.  As shown in
ref.~\cite{MMRT}, in the chiral limit, there is a unique choice of the
coefficients of the operators which belong to different chiral
representations, i.e. $C_i^{(\pm)}$, $C_\sigma^{(\pm)}$,
$C_s^{(\pm)}$ and $C_p^{(\pm)}$, such that the subtracted operators
$O^{(\pm)}_{\mathrm sub}$ satisfy this Ward identity.
By varying the points $x_1, x_2, \cdots, x_n$, eq.~(\ref{eq:giwi})
corresponds to an overdetermined set of linear inhomogeneous equations
which, in principle, allow for the determination of all the mixing
coefficients. Using the property that 
the axial current is conserved in the chiral limit, 
it is convenient to rewrite eq.~(\ref{eq:giwi}) as follows:
\begin{equation}
\sum_{\vec y} \langle \left(A^f_4(\vec y, t_b) -  
A^f_4(\vec y, -t_a)\right)
O^{(\pm)}_{\mathrm sub}(0)\,\Phi(x_1, x_2, \cdots, x_n)\rangle = -i
\langle\frac{\delta O^{(\pm)}_{\mathrm sub}(0)}{\delta\alpha^f}
\,\Phi(x_1, x_2, \cdots, x_n)\rangle \ .
\label{eq:giwi2}\end{equation}
The above equation shows that there are no contact terms arising upon
integration over $y$~\cite{luscher}. The absence of contact terms
implies that we do not have to include the mixing with operators that
vanish by the equations of motion. Although these operators do not
contribute to physical matrix elements, in general (see below) they
must be taken into account in the determination of the mixing
coefficients. Notice that the coefficients determined in the chiral limit
are sufficient to predict unambiguously the physical $K\to\pi\pi$
amplitudes. 

The other two non-perturbative methods mentioned above use
quark and gluon correlators in a fixed gauge, and either impose
normalization conditions at large Euclidean momenta \cite{talevi}, or
enforce the Ward Identities on quark Green functions~\cite{jlqcd}.
The problem with these methods is that they require the inclusion of
two additional classes of operator\footnote{ For a more detailed
discussion of why such operators appear see ref.~\cite{dawson}.}:
\begin{enumerate}
\item
Gauge invariant operators which vanish by the equation of motion.
These do not contribute to on-shell matrix elements,
but do contribute to the off-shell correlators used in these methods.
In the present case there are three such operators of low enough dimension 
\bea
&&\bar s (\rDslash + m_d) d + \bar s (-\lDslash + m_s) d
\,,\nonumber\\
&&
\bar s (\rDslash + m_d)^2 d + \bar s (-\lDslash + m_s)^2 d 
\,,\\
&&
\bar s (-\lDslash + m_s) (\rDslash + m_d) d
\,.\nonumber
\eea
The coefficient of the first is proportional to $(m_c-m_u)/a$,
while that of the latter two are proportional to $(m_c-m_u)$.
These operators also appear in the continuum, but they can
be removed in perturbation theory by studying how the form factors behave
as one goes on-shell. This is not possible in a numerical simulation.
\item
Operators which are not gauge invariant.
These are, however, constrained to be either BRST invariant,
or to vanish by the equations of motion \cite{zuber}.
Note that there is an exact BRST symmetry on the lattice after 
gauge fixing \cite{gaugefix}.
There are two such operators of low enough dimension, and with
positive CPS parity,
\bea
&&(m_c-m_u) \left[\bar s \ldslash (\rDslash + m_d) d - 
\bar s (-\lDslash + m_s) \rdslash d\right] \,,\\
&&(m_c-m_u) \left[\bar s \rdslash (\rDslash + m_d) d - 
\bar s (-\lDslash + m_s) \ldslash d\right] 
\,.
\eea
\end{enumerate}
By varying the external momenta, and using suitable
projectors~\cite{talevi,jlqcd}, it may be possible, in principle, to
separate the contributions of these operators.  The inclusion of five
additional operators, when the normalization conditions are imposed on
quark states in momentum space, seems, however, to make the
possibility of a non-perturbative determination of the mixing
coefficients quite remote in practice. 

These problems can be avoided, by working in configuration space,
rather than in momentum space. This can be achieved by studying
(unamputated) quark Green functions of the form
\begin{equation}
\langle \psi_1(x_1)\psi_2(x_2)O^{(\pm)}(0)\bar\psi_3(x_3)\bar\psi_4(x_4)
\rangle\ ,
\end{equation}
where all the points are separated. In this case none of the operators
which vanish by the equation of motion can appear. To enforce the Ward
identities, one uses eq.~({\ref{eq:giwi2}) with a non-gauge
invariant $\Phi(x_1,\cdots,x_4) = 
\psi_1(x_1)\psi_2(x_2)\bar\psi_3(x_3)\bar\psi_4(x_4)$. The expectation 
value has to be evaluated in a fixed gauge. Although, in this case,
we have the same number of coefficients to determine, we prefer the 
GIWI method, since it requires the evaluation of gauge invariant
correlation functions only.

\section{Construction of ${\mathbf{\cal H}^{W}_{\rm eff}}$
from first principles}
\label{sec:jj}

In this section we propose a method which, in principle, avoids
the difficulties caused by mixing with lower dimension operators,
and which automatically gives the effective weak Hamiltonian with the 
correct normalization.
In addition, it allows one to construct an improved weak Hamiltonian,
i.e. one having errors of $O(a^2)$,
given only the improved versions of the weak currents.
The method does not use chiral perturbation theory, and thus applies
equally well to the $\Delta S=1$, $\Delta C=1$ and $\Delta B=1$ parts of
the weak Hamiltonian. 
The method is speculative in the sense that it is likely to require
more computational power than is presently available,
although we expect it to be practical with the advent of Teraflops machines.

The standard construction of the non-leptonic weak Hamiltonian
begins with the  expression 
\begin{equation}
{\cal H}_{\rm eff}^{W}=g^2_{W}\int d^4x\,D_{\rho\nu}^W(x;M_{_W}) 
T\left[J_{\rho L}(x) J^\dagger_{\nu L}(0)\right] \, ,
\label{eq:HEFF}
\end{equation}
where
\begin{equation}
D_{\rho\nu}^{W}(x;M_{_W})=\int d^4p\,\frac{\mbox{e}^{ipx}}
{p^2+M_{_W}^2} (\delta^{\rho\nu}-\frac{p^\rho p^\nu}{M_{_W}^2})
\label{WPROP}
\end{equation}
is the $W$-boson propagator and $J_{\rho L}$ is the (left-handed)
hadronic weak current.  One then performs an operator product
expansion (OPE) on the product of the two currents in
eq.~(\ref{eq:HEFF}), which is justified by the observation that the
dominant contribution to the integral comes from distances $|x| \ll
M_{_W}^{-1}$. For physical amplitudes, one obtains in this way
\begin{equation}
\langle h|{\cal H}_{\rm eff}^{W}|h'\rangle =
\frac{G_{F}}{\sqrt{2}} \sum_i C_i(\mu,M_{_W}) M_{_W}^{6-d_i} 
\langle h|{O}^{(i)}(\mu)|h'\rangle\ ,
\label{eq:HEFFOPE}
\end{equation}
where $d_i$ is the dimension of the operator ${O}^{(i)}(\mu)$,
and functions $C_i(\mu,M_{_W})$ result from the integration
of the Wilson expansion coefficients, $c_i(x;\mu)$ (defined in 
eq.~(\ref{eq:ME}) below), with the
$W$-propagator. Schematically, suppressing Lorentz indices, one has
\begin{equation}
C_i(\mu,M_{_W}) M_{_W}^{6-d_i} 
= \int d^4x\,D^{W} (x;M_{_W}) c_i(x;\mu)\ .
\label{eq:WILCOEF}
\end{equation}
The ${O}^{(i)}(\mu)$ are quark and/or gluon operators
renormalized at the subtraction point $\mu$. 
The functions $C_i(\mu,M_{_W})$ are evaluated in perturbation theory and their
running with $\mu$ is dictated by the renormalization group equation
which follows from the $\mu$-independence of the l.h.s. of
eq.~(\ref{eq:HEFFOPE}). 

The sum in the expansion~(\ref{eq:HEFFOPE}) is over operators of
increasing dimension.  We consider in the following only operators
with dimensions $d_i \le 6$, since the contribution from operators
with $d_i>6$ is suppressed by powers of $1/M_{_W}$.

All the intricacies of operator mixing in the definition of the finite
and renormalized operators, ${O}^{(i)}(\mu)$, come about because the
integrals in~(\ref{eq:HEFF}) and~(\ref{eq:WILCOEF}) are extended down
to the region of extremely small $x$. The complicated mixing for the
${O}^{(i)}(\mu)$'s in terms of bare operators arises from contact
terms when the separation of the two currents goes to zero (i.e. when
$|x|$ is of the order of $a$). The problem is particularly bad because
chiral symmetry is broken by the lattice regularization.  This
observation suggests that a simple way to avoid these complications is
to implicitly define the renormalized operators by enforcing the
OPE on the lattice for distances $|x|$ much larger than the lattice
spacing $a$. We imagine proceeding in the following way:
\begin{enumerate}
\item 
Take the $T$-product of two properly normalized weak currents,
$J_{\rho L}(x) J^\dagger_{\rho L}(0)$. If required these currents can be improved.
\item
Measure the hadronic matrix element
$\< h\vert T[J_{\rho L}(x) J^\dagger_{\rho L}(0)]\vert h'\>$
in a Monte Carlo simulation,
as a function of $x$ for $|x|\rightarrow 0$ in the region 
\begin{equation}
a\ll |x|  \ll  \Lambda_{QCD}^{-1} 
\,.\label{eq:COND}
\end{equation}
\item
Extract the numbers $\< h\vert {O}^{(i)}(\mu)\vert h'\>$ by
fitting in the region~(\ref{eq:COND}) the $x$-behaviour of 
$\< h\vert T[J_{\rho L}(x) J^\dagger_{\rho L}(0)]\vert h'\>$ to the formula 
\begin{equation}
\< h\vert T\left[J_{\rho L}(x) J^\dagger_{\rho L}(0)\right]\vert h' \> = \sum_i
c_i(x;\mu) \< h\vert {O}^{(i)}(\mu)\vert h' \> 
\,,\label{eq:ME}
\end{equation}
where the Wilson coefficients $c_i(x;\mu)$ are determined by
continuum perturbation theory using any standard renormalization scheme.
The scale $\mu$ should be chosen so that $1/\mu$ lies in the
range defined by eq.~(\ref{eq:COND}). Since 
we only consider operators of dimension 6 or lower, the $T$-product differs
from the right-hand side of eq.~(\ref{eq:ME}) by terms of $O(|x|^2\lqcd^2)$,
which is an estimate of the size of the systematic errors in this procedure.
Note that in eq.~(\ref{eq:ME}) an average over the points $x$ and $-x$
(schematically $J(x)J(0)\to 1/2(J(x)J(0) + J(-x)J(0))$)
is implied in order to eliminate from the OPE terms which do not appear
in physical amplitudes because of the integration over $x$ in 
eq.~(\ref{eq:HEFF}). These terms, however, would appear on the 
r.h.s. of eq.~(\ref{eq:ME}) were we to not perform this average.

\item
Insert the numbers $\< h\vert {O}^{(i)}(\mu)\vert h'\>$ 
determined in this way into the expression for the 
matrix elements of  ${\cal H}_{\rm eff}^{W}$, finally obtaining 
eq.~(\ref{eq:HEFFOPE}).
\end{enumerate}

For the implementation of this procedure, what is required is the
existence of a window, eq.~(\ref{eq:COND}), in which the distance
between the two currents is small enough so that perturbation theory
can be used to determine the expected form of the OPE, but large
enough that lattice artifacts are small. These artifacts will be
suppressed by powers of $a/x$.  Clearly the existence of such a window
requires that we have a sufficiently small lattice spacing. At the
same time the physical volume of the lattice must be sufficiently
large to allow the formation of hadrons.

A few remarks may be useful at this point:
\begin{itemize}
\item
The method determines directly the ``physical'' 
matrix elements of the operators appearing in the OPE of the 
two currents, i.e. the matrix elements of the finite, renormalized operators
${O}^{(i)}(\mu)$, without any reference to the magnitude of the $W$-mass. 
Thus we do not need to probe distances of $O(1/M_{_W})$ with lattice
calculations.
\item
If the action and currents are improved, 
then the resulting matrix elements of ${O}^{(i)}(\mu)$,
and thus of ${\cal H}_{\rm eff}^{W}$, will also be improved.
\item 
The $\mu$-dependence of the matrix elements of the operators
${O}^{(i)}(\mu)$ is given trivially by that of the (perturbative)
Wilson coefficients, $c_i(x;\mu)$.  It compensates the related
$\mu$-dependence of the functions $C_i(\mu,M_{_W})$ in such a way that
the l.h.s of eq.~(\ref{eq:HEFFOPE}) is independent of the choice of the
subtraction point. A similar comment holds for the dependence on
renormalization scheme.
\item 
Unlike the methods discussed in previous sections,
this approach automatically yields hadronic amplitudes that are properly 
normalized (in the renormalization scheme in which the 
Wilson coefficients appearing in eq.~(\ref{eq:ME}) are computed).
\end{itemize}

We now discuss the feasibility of the method in more detail.
The critical step is fitting to the form predicted by the OPE,
eq.~(\ref{eq:ME}).
Typically more than one operator contributes to the sum,
so one must be able to separate the contributions using their
different dependence on $x$.
The operators of interest are of dimension 6,
and thus have Wilson coefficients which vary logarithmically with $x$.
At leading order the form is
\be
c_i(x;\mu) \propto 
\left(\alpha_s(1/x) \over \alpha_s(\mu)\right)^{\gamma_0^{(i)}\over 2 \beta_0}
= 1 + \frac{\alpha_s}{4\pi}\gamma_0^{(i)} \log(x \mu) + \dots
\,,
\label{eq:formofci}
\ee
where $\gamma_0^{(i)}$ is the one-loop anomalous dimension of the operator
$O^{(i)}$, and $\beta_0$ is coefficient of the one-loop term in the
$\beta$-function.
By contrast, the coefficients of lower dimension operators diverge
as powers of $1/x$ (up to logarithmic corrections).
Thus if lower dimension operators are present they will dominate at
short distances, and it will be very difficult to pick out the matrix
elements of the dimension 6 operators.
If, on the other hand, only dimension 6 operators appear then
it may be possible to separately determine their matrix elements.
How feasible this is depends on how large a range of $x$ we can use,
and on the magnitude of the differences between the anomalous dimensions.

Fortunately, in the cases of interest, there are no operators of
dimension lower than 6.  Consider, for example, the $\Delta S=1$ part
of ${\cal H}_{\rm eff}^{W}$.  The operators which can appear in the
OPE are $\oplmi$ [defined in eq.~(\ref{eq:oplmidef})], and in addition
\be
O' = (m_c^2-m_u^2) \,
\bar s (\overrightarrow{D_\mu} - \overleftarrow{D_\mu}) \gamma_\mu^L d
\,.
\ee
The GIM mechanism requires $O'$ to vanish when $m_c=m_u$, while chiral
symmetry requires both the quarks to be left-handed and that the GIM
factor be quadratic in the quark masses. Although this operator looks
new, its negative parity part is, by the equations of motion,
proportional to $(m_c^2-m_u^2)\,O_p$, where $O_p$ is defined in
eq.~(\ref{eq:opdef}), and its positive parity part is proportional to
$(m_c^2-m_u^2)\, (m_s + m_d)\, O_s$. So these are the same operators
we encountered in sections~\ref{sec:problem} and \ref{sec:pp}, except
for the overall factors.  Since $O'$ has dimension 6,, its coefficient
function depends only logarithmically on $x$.\par

To determine the matrix elements using eq.~(\ref{eq:ME}) we need the
anomalous dimensions, which for the three operators are
\be
\gamma^{(+)}_0 = 4, \qquad 
\gamma^{(-)}_0 = -8, \qquad
\gamma '_0 = 16\ .
\label{eq:ANOMALDIM}
\ee
In fact, the contribution of $O'$ to the r.h.s. of eq.~(\ref{eq:ME})
can be determined separately since its matrix element does  
not require any
subtraction and can be calculated directly.  As for $O^{\pm}$, since
their anomalous dimensions are well separated from one another, it may
be possible to extract the corresponding matrix elements and then
construct the physical amplitude of ${\cal H}_{\rm eff}^{\Delta S=1}$.

An important element of the procedure proposed in this section is that
since it is the continuum OPE which determines the operators which appear,
these are restricted by continuum symmetries.  This is
because, for $|x|\gg a$, the lattice OPE matches
that of the continuum with discretization errors suppressed by
powers of $a/x$.  

The previous discussion shows how we can, in principle, remove the
contribution of the scalar and pseudoscalar densities from any
hadronic matrix element.  Not only does the method work for both
positive and negative parity parts of ${\cal H}_{\rm eff}^{W}$, but it
also works if the weak Hamiltonian carries momentum $\Delta p$.  Thus
the simplest way to test the method may be to calculate $K\to\pi$
matrix elements and then use chiral perturbation theory to relate
these to the physical amplitude, as described in the previous section.
The problems of operator mixing described in sec.~\ref{sec:pp} do not
apply to the new method.

Of course, for physical matrix elements 
one does not need to worry about the subtraction of $O'$.
This is because the matrix elements determined by this method
are those of the continuum operator, up to discretization errors.

We end this section with an observation on the computational
feasibility of this approach.  The main difficulty is to have a
sufficiently large range of values of $|x|$ in order to separate the
contributions from the different operators, and yet to satisfy the
condition~(\ref{eq:COND}). These constraints make the method difficult
at present, and it will only be fully exploited when Teraflops machines
become available.

\section{A propagating top quark}
\label{sec:top}

For CP violating processes in kaon decays, or for $B$ decays where
top-penguin diagrams enter at a Cabibbo-allowed level, the strategies
described in sec.~\ref{sec:new} and \ref{sec:jj} for the negative
parity operators fail, because the GIM mechanism is not operative.  In
particular $O^{(\pm)}$ mix with all the penguin operators (see
below). This makes the calculation of the mixing matrix of comparable
difficulty to that for the positive parity operators described in
sec.~\ref{sec:pp}. For positive parity operators, the analysis of
section~\ref{sec:pp} still applies. The difference is that the mixing
coefficients of the magnetic operator and scalar density become more
divergent~\cite{MMRT} and this can make the numerical determination of
the renormalized operators less precise.  In order to circumvent these
problems we propose two methods involving a fictitious top quark (with
mass $\tilde m_t$) which is light enough to propagate on the lattice.

The basic idea is to work with two different scales: the first, $\mu$,
is larger than $\tilde m_t$, so that the corresponding operator basis
is as in the previous sections ($O^{(\pm)}$); the second,
$\mu^\prime$, is smaller than $\tilde m_t$ so that a full set of
penguin operators is generated.  The matrix elements of the operators
renormalized at the scale $\mu$ are computed numerically following the
strategies explained in secs.~\ref{sec:new}--\ref{sec:jj}. By matching
the result to the amplitude expressed in terms of operators
renormalized at $\mu'$, we extract their matrix elements. In this way,
at least in principle, we can obtain the matrix elements of the
penguin operators without directly computing them. We now present
the details of this procedure.

At scales $\mu'$ below $m_t$, when GIM is not operative, the form of
the $\Delta S=1$ effective Hamiltonian is
\bea
{\cal H}_{eff}^{\Delta S=1}&=&\frac {G_F} {\sqrt{2}}
\Bigl[ \lambda_u\, \Bigl( C_1(\mu^\prime,M_{_W})\left( 
Q^u_1(\mu^\prime) - Q_1(\mu^\prime) \right) +
C_2(\mu^\prime,M_{_W})
\left( Q^u_2(\mu^\prime) - Q_2(\mu^\prime) \right)  \Bigr)\nn\\
&-&\lambda_t \, \vec C(\mu^\prime,M_{_W}, {m}_t)
\cdot\vec Q(\mu^\prime)   \Bigr]
\protect\label{eq:ehmu}
\eea
where $\lambda_q=V_{qd} V_{qs}^*$, $q=u,c,t$ ($\lambda_c$ is
eliminated by using the unitarity relation $\lambda_c =
-\lambda_u-\lambda_t$). Here $\vec Q$
contains the ``penguin'' operators
\begin{equation}
\vec Q(\mu^\prime)\equiv
\left(Q_1(\mu^\prime),Q_2(\mu^\prime),\dots, Q_{6}(\mu^\prime)\right)
\end{equation}
and $\vec C$ are the corresponding coefficients
\begin{equation}
\vec C(\mu^\prime)\equiv(
C_1(\mu^\prime,M_{_W}),C_2(\mu^\prime,M_{_W} ), 
C_3(\mu^\prime,M_{_W},{m}_t) \dots, 
C_{6}(\mu^\prime,M_{_W}, {m}_t))
\,.
\end{equation}
A convenient basis of operators
when $QCD$ corrections are taken into account is~\cite{russi}--\cite{lus}
\bea
Q_{ 1}&=&({\bar s}d)_{ (V-A)}
    ({\bar c}c)_{ (V-A)}\,, 
\nonumber\\
Q_{ 2}&=&({\bar s}c)_{ (V-A)}
    ({\bar c}d)_{ (V-A)}\,, 
\nonumber\\
Q_{ 3,5} &=& ({\bar s}d)_{ (V-A)}
    \sum_{q}({\bar q}q)_{ (V\mp A)}
\protect\label{eq:basis}\\
Q_{ 4} &=&\sum_{q} ({\bar s}q)_{ (V-A)}
    ({\bar q}d)_{ (V - A)}
\nonumber\\
Q_{6} &=& -2 \sum_{q} ({\bar s}q)_{ (S+P)}
    ({\bar q}d)_{ (S-P)} \nonumber
\,.\eea
$Q_1^{u,t}$ and $Q_2^{u,t}$ are the analogous operators to $Q_1$ and
$Q_2$ with the up- and top-quark replacing the charmed one.
Here the subscripts $(V \pm A)$ and $(S \pm P)$ indicate the chiral
structures, and the sum over quarks $q$ runs over the active flavours
at the scale $\mu^\prime$.  For simplicity we ignore electroweak
penguin and magnetic operators; it is straightforward to generalize
the following discussion to include them.  The coefficient functions
appearing above have been calculated up to non-leading order in
perturbation theory in refs.~\cite{noi}--\cite{ciuz}.  The effective
Hamiltonian relevant for $\Delta B=1$ decays is simply obtained by
replacing the $s$ quark with the $b$ quark.

The part of ${\cal H}_{\rm eff}^{\Delta S=1}$ proportional to $\lambda_u$
is the same as that considered above in eq.~(\ref{eq:HW}),
and has been the focus of discussion for much of the paper.
We have simply re-expressed it here in the new operator basis,
in terms of which 
\begin{eqnarray}
O^{\pm} &=& (Q_1^u - Q_1) \pm (Q_2^u - Q_2) \,, \\
C_\pm(\mu, M_{_W}) &=& \frac12
\left[ C_1(\mu, M_{_W}) \pm C_2(\mu, M_{_W}) \right] \,.
\end{eqnarray}
This part of ${\cal H}_{\rm eff}^{\Delta S=1}$
gives the dominant contribution to CP conserving $K\to\pi\pi$
amplitudes.
These amplitudes can be calculated using the methods presented in 
secs.~\ref{sec:new}--\ref{sec:jj}. 

The difficulties arise for the part of ${\cal H}_{\rm eff}^{\Delta S=1}$
proportional to $\lambda_t$, which gives rise to CP violation in kaon decays.
This part contains the penguin operators $Q_i$, whose matrix elements
are not protected by the GIM mechanism.
To write a renormalized version of these operators
requires subtracting $O_p$ and $\tilde O_\sigma$ with appropriate
coefficients, and accounting for mixing with all the other operators
$Q_j$. This is true for both positive and negative parity sectors.
The methods of this section are designed to avoid these difficulties.

We do so by introducing a dynamical top quark so as to keep
the GIM mechanism operative.
This will, however, be a fictitious
top quark with mass satisfying
\begin{equation}
1/a \gg \tilde{m}_t \gg m_c \gg \Lambda_{\rm QCD} \,.
\end{equation}
In other words, our top is light enough to propagate on the lattice,
but, like the physical top, it is heavier than the charm quark.  As
explained below, by using the fictitious top we can extract the matrix
elements $ \langle h \vert Q_i(\mu') \vert h^\prime \rangle$ which can
then be inserted into the expression for ${\cal H}_{\rm eff}^{\Delta S=1}$,
eq.~(\ref{eq:ehmu}).  In this respect the method is similar to that of
sec.~\ref{sec:jj}. For purposes of illustration we will restrict the
discussion below to negative parity operators.

We begin with the two matrix elements 
\be 
{\cal M}_i(\mu,\tilde m_t) \equiv \langle h \vert 
 Q_i(\mu) - Q^t_i(\mu) 
\vert h^\prime \rangle \, ,
\quad i=1,2 
\label{eq:match1}\ee
evaluated at a renormalization scale satisfying $a^{-1} \sim \mu \gg
\tilde m_t$.  Since GIM is operative, the analysis of
sec.~\ref{sec:problem} applies (with $m_u\to m_t$).  Thus we can
define $Q_i(\mu) - Q^t_i(\mu)$ in terms of bare lattice operators by
\be
Q_i(\mu) - Q^t_i(\mu) = Z_{ij}(\mu a,g_0^2)
\left[ Q_j(a) - Q^t_j(a) + (m_c-\tilde{m}_t) {C_p^{(j)} \over a} O_p(a)
\right]
\,,
\ee 
where $i,j=1,2$ and  
the subtraction coefficients $C_p^{(j)}$ are determined by 
enforcing
\be
\langle 0 \vert
Q_j(a) - Q^t_j(a) + (m_c-\tilde{m}_t) {C_p^{(j)} \over a} O_p(a)
\vert K \rangle = 0
\,.
\ee
The $Z_{ij}$ are related by a simple change of basis to the
$Z^{(\pm)}$ of eq.~(\ref{eq:rops}), and can be calculated either
perturbatively or non-perturbatively.  In this way we can obtain
${\cal M}_i(\mu,\tilde m_t)$ from a lattice calculation, as a function
of $\tilde{m}_t$, for some choice of $\mu$.  A simple choice is $\mu
\approx 1/a$.

On the other hand, we can also consider the same matrix
elements for a renormalization scale
$\tilde{m}_t\gg \mu' \gg m_c$. In this case, the GIM mechanism is not
operative, and the matrix elements can be expressed in terms of the
six operators which appear in eq.~(\ref{eq:basis})
\be 
 {\cal M}_i(\mu,\tilde m_t) =
 \sum_{j=1,6} \hat Z^{-1}_{ij}(\mu^\prime,\mu, \tilde{m}_t) 
\langle h \vert Q_j(\mu^\prime) \vert h^\prime \rangle
\, .  \label{eq:match2} 
\ee
The rectangular matrix $\hat Z^{-1}$ can be calculated
perturbatively by matching the theory with and without the fictitious
top quark.
This is exactly the method used to calculate the coefficients
$\vec C$ in the theory with the physical top quark mass,
except that in the physical case one must simultaneously integrate
out both the $W$ boson and the top quark. Here we are effectively
integrating  out the $W$ first and then the fictitious top quark.
The results for $\hat Z^{-1}$ at non-leading order can be reconstructed 
from those computed 
in refs.~\cite{noi}--\cite{ciuz}.
The running in the $1$--$2$ submatrix is particularly simple (since the
operators $Q_{3}$--$Q_{6}$ do not feed back into $Q_1$ and $Q_2$)
and one can show that
\be 
Z_{ij}(\mu^\prime a , g_0^2)=
\sum_{k=1,2} \hat Z_{ik}(\mu^\prime, \mu )  Z_{kj}(\mu a ,  g_0^2) \, ,
 \label{eq:matchzz}\ee
with  $(i,j)=1,2$. Here $\hat Z_{ik}$ is the inverse of the $2\times 2$
sub-block of the mixing matrix appearing in eq.~(\ref{eq:match2}).
This submatrix does not depend on $\tilde m_t$, but only on 
$\mu'$ and $\mu$.

We now choose a value of $\mu^\prime$ and vary $\tilde m_t$.  The six
matrix elements of interest, $\langle h \vert Q_j(\mu^\prime)
\vert h^\prime \rangle$, are obtained by fitting the right hand side
of eq.~(\ref{eq:match2}) to ${\cal M}_i(\mu,\tilde m_t)$ computed
numerically as in eq.~(\ref{eq:match1}), and using the renormalization
matrix $\hat Z^{-1}$ calculated perturbatively. 
Since the dependence on $\tilde m_t$ is
logarithmic, this will not be easy.  The procedure is analogous to our
use of the $x$-dependence in sec.~\ref{sec:jj} to separate the
renormalized matrix elements of operators appearing in the weak
Hamiltonian. Having determined these renormalized matrix elements, we
can insert them into the expression (\ref{eq:ehmu}) for the effective
Hamiltonian.  At this point the constraint $M_{_W} \gg \tilde{m}_t$ can
be removed since the Wilson coefficients of the operators appearing in
${\cal H}_{eff}^{\Delta S=1}$ can be computed perturbatively for
arbitrary values of $\tilde m_t$, including $\tilde m_t=m_t$.

Before discussing the errors involved in this procedure, we make the
following observation. In the effective theory where the top quark has
been removed, provided that $\mu' \gg \lqcd$, we can evolve the
renormalized operators from one scale to another using perturbation
theory. In particular we can obtain $\langle h \vert Q_j(\mu) \vert
h^\prime \rangle$ from $\langle h \vert Q_j(\mu') \vert h^\prime
\rangle$ for all six operators using the $6\times 6$ anomalous
dimension matrix computed
in perturbation theory with the effective Hamiltonian where the top
quark has been integrated out. Thus we can directly extract the matrix
elements of the operators $\vec Q(\mu)$ from ${\cal M}_i(\mu, \tilde
m_t)$.  Note, however, that
an accurate determination of the matrix elements of
$\vec Q(\mu)$ (or of $\vec Q(\mu^\prime)$) requires that the typical scale,
$\Lambda_{hh'}$, of masses and external momenta appearing in the
physical process $h^\prime \to h$ be much smaller than $\tilde
m_t$.  This is because in the matching procedure we neglect terms of
$O(\Lambda_{hh'}/\tilde m_t)$. 

An alternative method, in the same spirit as the approach followed in
sec.~\ref{sec:jj}, is the following.  We can avoid the need for any
non-perturbative subtraction by separating the two currents and using
a fictitious propagating top quark.  Thus we directly match $\langle
h\vert T(J_{\rho L}(x) J^\dagger_{\rho L}(0)) \vert h'\rangle_{\rm
top}$, where the subscript indicates the presence of the fictitious
top, to the formula
\begin{equation}
\langle  h\vert T(J_{\rho L}(x) J^\dagger_{\rho L}(0))
\vert h' \rangle_{\rm top} = \sum_{i=1,6}
c_i(x; \mu^\prime, \tilde{m}_t) \langle h\vert
 {O}^{(i)}(\mu^\prime)\vert h' \rangle
\,,\label{eq:MEtop}
\end{equation}
where 
\bea J_{\rho L}(x)  J^\dagger_{\rho L}(0)&=& 
 \bar s(x) \gamma_\rho (1 -\gamma_5) t(x)
\bar t(0) \gamma_\rho (1 -\gamma_5) d(0) \nn \\ &-&
 \bar s(x) \gamma_\rho (1 -\gamma_5) c(x)
\bar c(0) \gamma_\rho (1 -\gamma_5) d(0) \, .\eea
The coefficients $c_{1,2}(x; \mu^\prime, \tilde{m}_t)\equiv
c_{1,2}(x; \mu^\prime)$ are the same as those in sec.~\ref{sec:jj}.  The
coefficients $c_i(x; \mu^\prime, \tilde{m}_t)$, with
$i=3$--$6$ are complicated functions of the anomalous dimension matrix
which can be worked out from the results of refs.~\cite{buras2} and
\cite{noi2} and computed numerically.

Both methods proposed in this section require small enough lattice
spacings to accommodate a number of scales. Like the method of
sec.~\ref{sec:jj} their full implementation is likely to require the
next generation of supercomputers.
 
\section{Conclusion}
\label{sec:concs}

In this paper we have suggested a number of new approaches with which
to study the \dirule\ using Wilson-like fermions. These methods can
also be used for staggered fermions.

In order to obtain the physical $K\to\pi\pi$ amplitude without relying
on chiral perturbation theory, or to study decays such as
$B\to\pi\pi$, one must learn how to extract information on final state
interactions from Euclidean amplitudes. The method of ref.~\cite{cfms}
might make this possible, but detailed numerical studies are needed to
assess whether it is practical.

The calculation of the CP violating part of $K\to\pi\pi$ amplitudes
with Wilson quarks is very difficult. A completely nonperturbative
method may require the addition of a fictitious top quark.

We have also reevaluated the method of ref.~\cite{MMRT} involving
$K\to\pi$ amplitudes. This approach is likely to be more difficult
because of the large number of mixing coefficients which have to be
determined non-perturbatively. It may however provide complementary
information to the results obtained with the $K\to\pi\pi$ method, 
and a check of the accuracy of chiral relations.

Only numerical studies will be able to confirm or refute our
present intuition that the $K\to\pi\pi$ methods are likely to provide
the better results in the near future.

\section*{Acknowledgements}
We would like to thank R. Gupta and A. Vladikas 
for discussions.  C.D., G.M., 
G.C.R., C.T.S., M.Ta. and M.Te. acknowledge partial support from EU
contract CHRX-CT92-0051.  M.Ta. thanks the Universit\`a di Roma ``La
Sapienza'' where part of the work for this paper was carried out and
acknowledges INFN for partial support and EPSRC for its support
through grant GR/K41663.  C.D., M.Ta. and C.T.S. acknowledge the PPARC
for its support through studentship 9530247X and grants GR/L22744 and
GR/J21569, respectively.  G.M., G,C.R., M.T. and M.T. acknowledge
partial support by M.U.R.S.T.  C.D. and S.S. thank the University of
Rome ``La Sapienza" for its hospitality, and the INFN for partial
support. S.S. was also partially supported by the U.S. Department of
Energy grant DE-FG03-96ER40956.

\end{document}